 \documentclass[final,5p,times,twocolumn]{elsarticle}

\usepackage{graphics}
\usepackage{tabto}
\usepackage{bm}
\usepackage{amsmath}
\usepackage{wrapfig,lipsum,booktabs}
\usepackage{amssymb}
\usepackage{url}
\usepackage{color}
\usepackage{titlesec}
\usepackage{mathtools, cuted}
\usepackage{widetext}
\newcommand{\fe}{\mathbf{\tilde{E}}}
\newcommand{\fb}{\mathbf{\tilde{B}}}
\newcommand{\fj}{\mathbf{\tilde{J}}}

\newcommand{\fk}{\mathbf{k}}
\newcommand{\fkhat}{\mathbf{\hat{k}}}
\newcommand{\parallelsum}{\mathbin{\!/\mkern-5mu/\!}}

\newcounter{bla}

\journal{Computer Physics Communications}
\begin{document}
\begin{frontmatter}

\title{Ultrahigh-order Maxwell solver with extreme scalability\\ for electromagnetic PIC simulations of plasmas.}

\author[1,2]{Henri Vincenti\thanks{henri.vincenti@cea.fr}\corref{author}}
\author[2]{Jean-Luc Vay\thanks{jlvay@lbl.gov}}

\cortext[author] {Corresponding author.\\\textit{E-mail address:} henri.vincenti@cea.fr}
\address[1]{LIDYL, CEA, CNRS, Universit\'e Paris-Saclay, CEA Saclay, 91 191 Gif-sur-Yvette, France}
\address[2]{Lawrence Berkeley National Laboratory, Berkeley, CA, USA}


\begin{abstract}
The advent of massively parallel supercomputers, with their distributed-memory technology using many processing units, has favored the development of highly-scalable local low-order solvers at the expense of harder-to-scale global very high-order spectral methods. Indeed, FFT-based methods, which were very popular on shared memory computers, have been largely replaced by finite-difference (FD) methods for the solution of many problems, including plasmas simulations with electromagnetic Particle-In-Cell methods. For some problems, such as the modeling of so-called ``plasma mirrors'' for the generation of high-energy particles and ultra-short radiations, we have shown that the inaccuracies of standard FD-based PIC methods prevent the modeling on present supercomputers at sufficient accuracy. We demonstrate here that a new method, based on the use of local FFTs, enables ultrahigh-order accuracy with unprecedented scalability, and thus for the first time the accurate modeling of plasma mirrors in 3D.
\end{abstract}

\begin{keyword}
Electromagnetic Particle-In-Cell method; Massively parallel pseudo-spectral solvers; Relativistic plasma mirrors; Pseudo-Spectral Analytical Time Domain solver; Finite-Difference Time-Domain solver
\end{keyword}

\end{frontmatter}

\section{Introduction} 
\subsection{Challenges in the modeling of Ultra-High Intensity (UHI) physics}
\begin{figure*}[h]
\centering
\includegraphics[width=\linewidth]{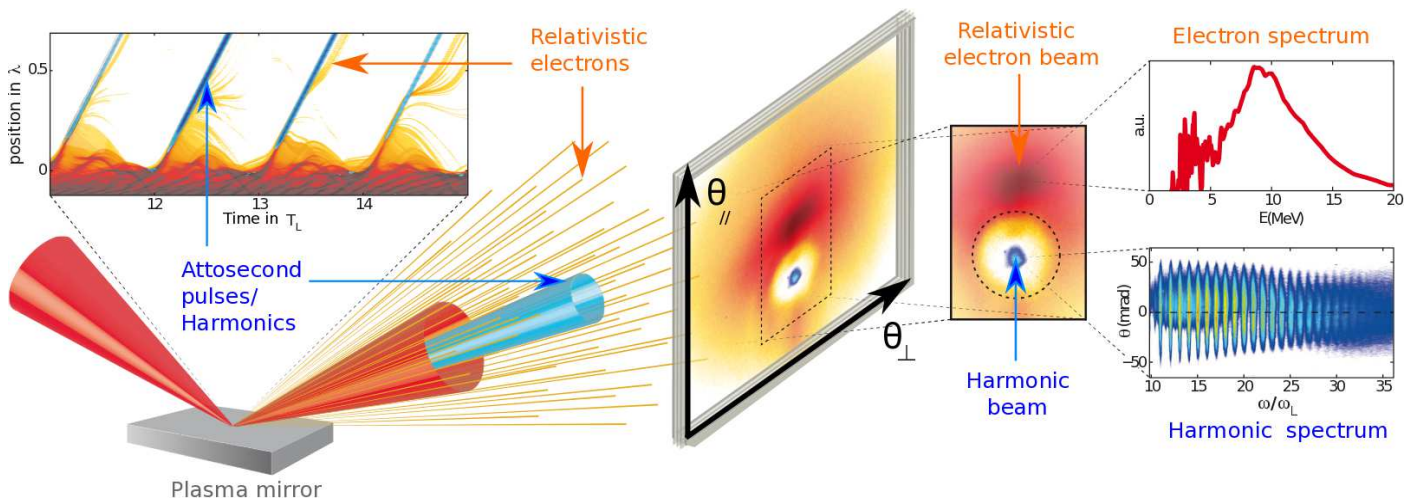}
\caption{\label{fig:UHI} \small Physics of UHI laser plasma mirror interaction. When exposed to UHI laser fields, plasma mirrors specularly
reflect the laser with a reflectivity higher than 70\% (bottom left image). Their highly non-linear coupling with the laser driving fields can lead to the emission of synchronized beams of high-order harmonics by Doppler effect \cite{thaury2007,PhysRevLett.99.085001} and relativistic attosecond electron bursts \cite{Thevenet2015,thevenet2016} (see upper left panel). Upper left panel was obtained from a 1D Particle-In-Cell simulation ran in the Bourdier's frame \cite{Bourdier1983}. Red color scale corresponds to plasma electron density and blue color scale to the harmonic field obtained after filtering off the laser frequency from the reflected field.  Typical spatial patterns of these beams obtained from experiments on UHI100 laser in CEA Saclay \cite{Thevenet2015}, are shown in the central images, and their typical energy spectra in the rightmost images. $\lambda$, $T_L$ and $\omega_L$ respectively stand for laser wavelength, laser period and laser frequency.}
\end{figure*}

The advent of high power petawatt (PW) femtosecond lasers has paved the way to a new, promising but still largely unexplored branch of physics called Ultra-High Intensity (UHI) physics \cite{RevModPhys.78.309}. Once such a laser is focused on a solid target, the laser intensity can reach values as large as $10^{22}W.cm^{-2}$, for which matter is fully ionized and turns into a ``plasma mirror'' that reflects the incident light \cite{PhysRevE.69.026402, thaury2007} (See Fig. \ref{fig:UHI}).  

The corresponding laser electric field at focus is so high, that ``plasma mirror'' particles (electrons and ions) get accelerated to relativistic velocities upon reflection of the laser on its surface. A whole range of compact ``tabletop'' sources of high-energy particles (electrons, protons, highly charged ions) and radiations ranging from X-rays to $\gamma$-rays may thus be produced from the interaction between this plasma mirror and the ultra-intense laser field at focus \cite{thaury2007,PhysRevLett.99.085001,Thevenet2015,Hegelich2006}.

The success of PW laser facilities presently under construction worldwide, which aim at understanding and controlling these promising particle and light sources for future application experiments \cite{Vincenti2014,PhysRevLett.108.113904,Wheeler2012}, will rely on the strong coupling between experiments and large-scale simulations with Particle-In-Cell (PIC) codes. 
Nevertheless, standard PIC codes currently in use partly fail to accurately describe most of UHI laser-plasma interaction regimes because the finite-difference time domain (FDTD) Maxwell solver produces strong instabilities and noise when the accelerated particles move at relativistic velocities \cite{Godfreyjcp74,Greenwoodjcp04} or when the produced short-wavelength radiations span broad emission angles and frequencies \cite{Blaclard2016}. With standard PIC codes, the mitigation of these instabilities often requires spatial and temporal resolutions that are so high that they are not practical for realistic 3D modeling on current petascale supercomputers and, it is projected, even on upcoming exascale machines. 

\subsection{Goal and outline of the paper}

To address this challenge, the solution that we propose here is to use highly precise pseudo-spectral methods to solve Maxwell's equations. Despite their high accuracy,  legacy pseudo-spectral methods employing global Fast Fourier Transforms (FFT) on the whole simulation domain have hardly been used so far in large-scale 2D/3D simulations due to their difficulty to efficiently scale beyond 10,000s of cores \cite{Gonoskov681092,Habibarxiv2012}, which is not enough to take advantage of the largest supercomputers required for 3D modeling.

 To break this barrier a pioneering grid decomposition technique was recently proposed for pseudo-spectral FFT-based electromagnetic solvers \cite{VayJCP2013}. The new technique was first validated by an extensive analytical work \cite{Vincenti2016a} and then implemented in our PIC code Warp+PXR. 
 
 
 In this paper, we will first demonstrate that the new technique enables, for the first time, the scaling of pseudo-spectral solvers on up to a million cores. We will then compare the speedup brought by our pseudo-spectral solvers against FDTD solvers in terms of time-to-solution, on a 3D simulation of relativistic plasma mirrors. 
 
The paper is divided in 4 sections: 

\begin{itemize}
\item In section 2: we briefly present the standard PIC method and detail its limitations in the modeling of UHI laser-plasma interactions, 
\item In section 3: we describe the new parallelization technique of pseudo-spectral solvers that we implemented in Warp+PXR and that enabled their scaling on up to a million cores, 
\item In section 4: we present scaling tests of our new implementation on the MIRA cluster at Argonne National Laboratory and the Cori cluster at the National Energy Research Scientific Computing Center in Berkeley. We also present performance benefits in terms of time-to-solution of the new solvers against FDTD solvers in the 3D modeling of plasma mirrors.  
\item In section 5: we present future implications of this work on UHI physics and beyond. 
\end{itemize}

\section{Limits of the standard Particle-In-Cell method} 


The electromagnetic Particle-In-Cell (PIC) method follows the evolution of a collection of charged macro-particles that evolve self-consistently with their electromagnetic fields. The core algorithm involves four operations at each time step: 1) evolve the velocity and position of the particles using the Newton-Lorentz equations, 2) deposit the charge and/or current densities through interpolation from the particles distributions onto the grid, 3) evolve Maxwell's electromagnetic wave equations on the grid, 4) interpolate the fields from the grid onto the particles for the next particle push. The most popular algorithm for solving Maxwell's wave equations is the Finite-Difference Time-Domain (or FDTD) solver: 
\begin{subequations}
\begin{eqnarray}
D_{t}\mathbf{B} & = & -\nabla\times\mathbf{E}\label{Eq:Faraday-2}\\
D_{t}\mathbf{E} & = & \nabla\times\mathbf{B}-\mathbf{J}\label{Eq:Ampere-2}
\end{eqnarray}
\end{subequations}
where the spatial differential operator is defined as $\nabla=D_{x}\mathbf{\hat{x}}+D_{y}\mathbf{\hat{y}}+D_{z}\mathbf{\hat{z}}$
and the finite-difference operators in time and space are defined
respectively as $ $$D_{t}G|_{i,j,k}^{n}=\left(G|_{i,j,k}^{n+1/2}-G|_{i,j,k}^{n-1/2}\right)/\Delta t$$ $
and $D_{x}G|_{i,j,k}^{n}=\left(G|_{i+1/2,j,k}^{n}-G|_{i-1/2,j,k}^{n}\right)/\Delta x$,
where $\Delta t$ and $\Delta x$ are respectively the time step and
the grid cell size along $x$, $n$ is the time index and $i$, $j$
and $k$ are the spatial indices along $x$, $y$ and $z$ respectively.
The difference operators along $y$ and $z$ are obtained by circular
permutation.

Even at relatively high resolution in space and time, the finite difference (FDTD) solver (used in the standard PIC formulation to integrate Maxwell's equations \cite{Birdsalllangdon}) can generate strong non-physical instabilities, which would affect the physics at play. One of the most popular FDTD solver uses the Yee scheme \cite{Yee}, which places fields on a staggered grid, giving second-order accuracy in space and time. Variations include non-standard FDTD schemes that average in the direction orthogonal to the stencils' derivative, for added benefits \cite{Karkicap06,CowanPRSTAB13} (labeled FDTD-CK in the section $4$). Those FDTD solvers use only spatially local information and must hence only exchange a few cells at the margin between each processor's assigned domain neighbors (guard cells, cf. Fig \ref{fig:data_exchange} (a)), thus achieving efficient parallelization up to millions of cores, as required for the simulation of large-scale problems. Nevertheless, it can produce significant unphysical degradation from discretization errors that are highly detrimental in the simulation of relativistic laser-plasma interactions. 

Using finite-difference solvers, the practically achievable level-of-accuracy is indeed strongly limited by numerical dispersion \cite{Blaclard2016} and numerical heating or noise, which are particularly critical for laser-plasma accelerator experiments where small unphysical errors can spoil the required high beam quality, or for simulations where accurate description of a large band of frequencies is required (e.g Doppler harmonics generated on relativistic plasma mirrors). The staggering of the electromagnetic field components on the Yee mesh also leads to errors due to inaccurate cancellation of self electric and magnetic fields components with charged particles moving at relativistic velocities \cite{Vaypop2008}. 

For a large majority of application experiments (e.g plasma harmonic generation spanning hundreds of harmonic orders) where a very accurate description of electromagnetic waves is required on a very large band of frequencies and angles, the resolution needed with finite difference solvers to accurately describe the physics would be so high that it would not be practical to perform a realistic 3D modeling on existing petascale supercomputers and, it is projected, not even on upcoming exascale machines \cite{Blaclard2016}. 

\section{New technique to build massively parallel pseudo-spectral PIC codes} 

To address this challenge, our solution is to use ultrahigh-order ($p$) solvers (up-to the infinite order limit $p\rightarrow\infty$) pseudo-spectral solvers to solve Maxwell's equations, which advance electromagnetic fields in Fourier space (rather than configuration space) and offer a number of advantages over standard FDTD solvers in terms of accuracy and stability.  

\subsection{Pseudo-spectral solvers for better accuracy}

In particular, Haber et al. \cite{Habericnsp73} showed that under weak assumptions, Fourier transforming Maxwell's equation in space yields an analytical solution for electromagnetic fields in time, called the Pseudo-Spectral Analytical Time Domain (PSATD) solver, which is accurate to machine precision for the electromagnetic modes resolved by the calculation grid. As a consequence, this solver enables infinite order ($p\rightarrow\infty$), imposes no Courant time step limit in vacuum and has no numerical dispersion. In addition, it represents naturally all field values at the nodes of a grid, thus eliminating errors associated with staggered field quantities. The PSATD algorithm advances the equations in Fourier ($\fk$) space as follows (see \cite{Habericnsp73} for the original formulation and \cite{VayJCP2013} for a more detailed derivation):

\begin{widetext}
\begin{subequations}
\begin{eqnarray}
\fe^{n+1} & = & C\fe^{n}+iS\fkhat\times\fb^{n}-\frac{S}{k}\fj^{n+1/2}
 + (1-C)\fkhat(\fkhat\cdot\fe^{n})\nonumber 
 +  \fkhat(\fkhat\cdot\fj^{n+1/2})\left(\frac{S}{k}-\Delta t\right),\label{Eq_PSATD_1}\\
\fb^{n+1} & = & C\fb^{n}-iS\fkhat\times\fe^{n}
+i\frac{1-C}{k}\fkhat\times\fj^{n+1/2}.\label{Eq_PSATD_2}
\end{eqnarray}
\end{subequations}
\end{widetext}

where $\tilde{a}$ is the Fourier Transform of the quantity $a$,  $\fkhat=\fk/k$, $C=\cos\left(k\Delta t\right)$ and $S=\sin\left(k\Delta t\right)$. As it turns out, the utilization of the wavenumbers $\fk$ in the PSATD algorithm corresponds to taking an infinite order approximation to the spatial derivative operators. In Fourier space, finite order $p$ approximations are obtained simply by substituting the formula $\mathbf{}k_{u}\rightarrow\sum_{j=1}^{p/2}w_{j}\frac{\sin\left(jk_{u}\Delta u\right)}{\Delta u/2}$ for the wavenumber component along the direction $u=\left\{ x,\, y\,\mathrm{or}\, z\right\} $ for centered finite differences on a nodal grid, where the $w_{j}$ are the coefficients of the finite-differenciation at order $p$ at position $j$ in real space. This substitution enables efficient Maxwell's solve at ultrahigh order (e.g. 100 or higher) much more efficiently than a finite-difference-based solver at the same order \cite{Gustafssonkreissoliger}, while significantly reducing the footprint of the stencil, as compared to infinite order \cite{Vincenti2016a}.

\subsection{New technique to scale pseudo-spectral solvers to a million cores and beyond}

\begin{figure*}[h]
\centering
\includegraphics[width=0.8\linewidth]{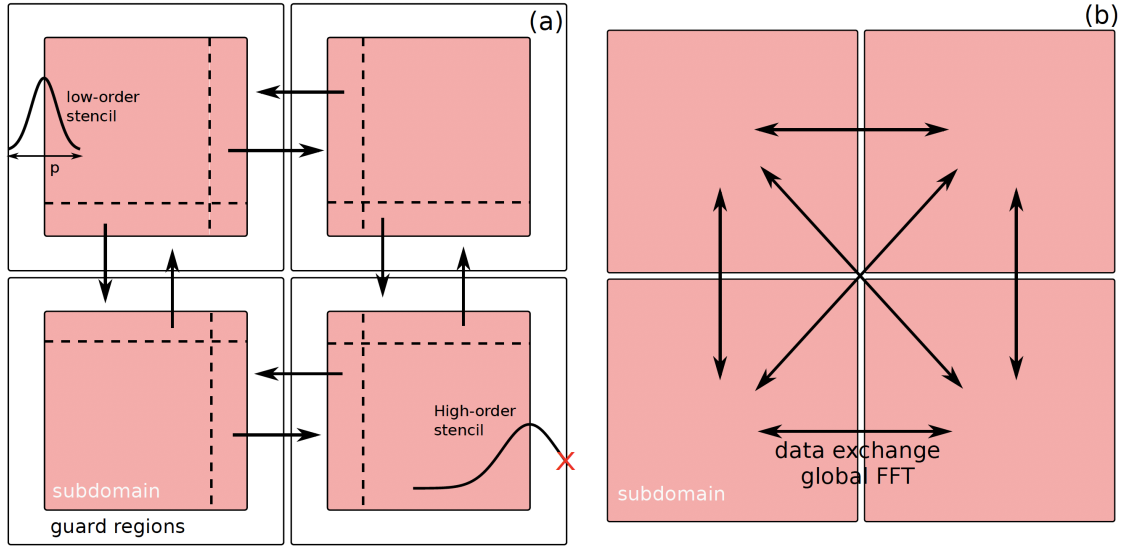}
\caption{\label{fig:data_exchange} \small (a) Standard Cartesian domain decomposition in 2D with exchange of guard cells used with most FDTD Maxwell solvers and with our new ultrahigh-order pseudo-spectral solver. Maxwell's equations are solved locally using local FFTs (b) Global 2D FFTs require global inter-processor communications during transposition of transformed arrays used in standard pseudo-spectral solvers.}
\end{figure*}

Nevertheless, despite significant advantages in terms of accuracy, ultrahigh-order pseudo-spectral solvers have not been widely adopted so far for large-scale simulations because of their poor scalability with increasing number of processors, which is due to the requirement of global inter-processor communications in the computation of global Fourier transforms (cf. Fig \ref{fig:data_exchange} (b)). Hence, while efficient strong scaling to millions of cores has been demonstrated with FDTD solvers, strong scaling with standard pseudo-spectral methods employing global FFTs have previously been reported only up to around 10,000s cores \cite{Gonoskov681092,Habibarxiv2012} and would thus not be adequate for realistic 3D modeling of UHI laser-plasma interactions using modern computers.

Recently, our team initiated a change of paradigm for simulation codes solving time-dependent problems where physical information propagates at a finite speed (e.g. Maxwell's equations). This new paradigm is based on using domain decomposition (standard for finite-difference solvers but not for spectral solvers) with spectral (FFT-based) solvers \cite{VayJCP2013}. This technique implies a small numerical approximation that falls off very rapidly with the number of guard cells surrounding each subdomain (as explained in more details below), allowing strong scaling of pseudo-spectral solvers to hundreds of thousands of cores and beyond. 

As in the case of low-order schemes, this technique divides the simulation domain into several subdomains (see Fig \ref{fig:data_exchange} (a)) with guard regions at their borders and Maxwell's equations are solved locally on each subdomain using local FFTs. 

\begin{figure}[h]
\includegraphics[width=\linewidth]{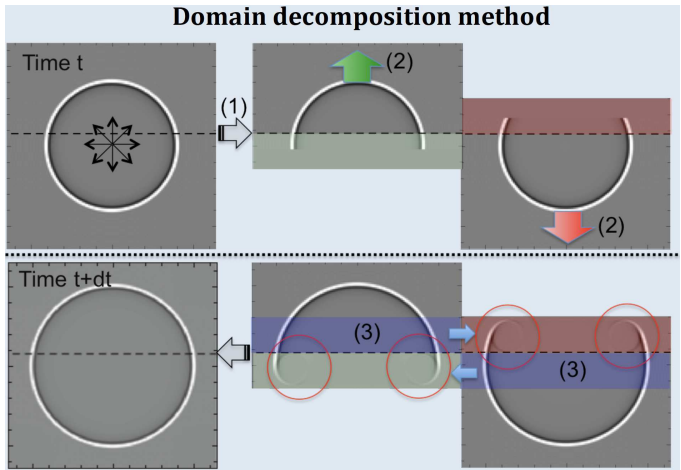}
\caption{\label{fig:test_1pulse} \small 2D propagation of a unit pulse with domain decomposition: (1) the field at time t is split into several subdomains (two here) with guard cells (green and red areas) (2) Fields in each subdomain are
advanced at time $t+\Delta t$ using PSATD solver (3) Guard cells at time $t+\Delta t$ containing spurious signal (inside red circles) are replaced between subdomains with unaffected field (blue areas)}
\end{figure}

For large order $p$ solvers and a finite number of guard cells $n_g$ such that $p/2>n_g$ or even $p/2\gg n_g$, Fig \ref{fig:data_exchange} (a) illustrates that this technique however implies some stencil truncations at subdomain boundaries that could generate spurious errors and that need to be characterized and controlled. The fundamental argument still legitimating this method is that physical information cannot travel faster than the speed of light (since we are solving Maxwell's equations). Choosing large enough guard regions should therefore ensure that spurious signal coming from these stencil truncations at subdomain boundaries would remain in guard regions and would not enter the simulation domain. This is illustrated in Fig. \ref{fig:test_1pulse} where a unit pulse is initialized at the center of a domain that has been decomposed into two subdomains (of unequal sizes). While some spurious signal is created in the guard cells' regions because of the domain truncation, it remains confined to the guard regions and does not enter the computational domain, thanks to the finite-speed of light. 


In 2016, we performed a comprehensive analytical and numerical study \cite{Vincenti2016a} that derived the exact expression of stencil truncation errors as a function of various numerical parameters (stencil order, number of guard cells, mesh resolution) and demonstrated that truncation errors are not growing even at very high orders $p$ and with a moderately low number of guard cells $n_g$. In particular, thanks to this study, we are now able to compute the number of guard cells required at a given order to have truncation error amplitudes lower than a given precision, including the zero machine precision. For instance, our model (validated by early numerical benchmarks) shows that 8 guard cells only are required at order 100 to get a negligible error that does not affect the UHI physics. 

This new paradigm opened the way to the use of these ultrahigh-order pseudo-spectral solvers at large scale for the accurate modeling of 3D laser-plasma interaction regimes that were previously out of reach of previous codes. Collaborators in Europe have also implemented the technique in a code co-developed with our team \cite{Jalas2017} (but not used in this study).

\section{Scaling tests and performance benefits of the new technique}
\subsection{How Performance Was Measured} 
Performances were measured on (a) full physic simulations of plasma mirrors and (b) Maxwell solver only. The performance simulations (b) were typically performed without I/O, but full scale physics simulations (a) with I/O were performed on up-to 270,000 cores. The full physics simulations (a) were performed on Argonne's National Laboratory Mira supercomputer as part of our 'PICSSAR' 2017 INCITE allocation, while simulations (b) were performed on Mira and U.S. DOE NERSC supercomputer Cori. This section presents the physical and numerical parameters used in simulations (a) as well as applications and kernels that were used for the simulations, the timing procedure, and the platforms Mira and Cori.

\subsubsection{Full physic simulations of plasma mirrors}

Recent experiments performed with the 100TW laser UHI 100 at CEA Saclay revealed a crucial feature of the emission from plasma mirrors \cite{Thevenet2015,thevenet2016}. Plasma mirrors act as injectors of attosecond electron bunches in the specularly reflected laser field that are further accelerated over distances of the Rayleigh length by Vacuum Laser Acceleration (VLA). The spatial pattern observed on the electron beam (see central panel on Fig \ref{fig:UHI}) shows a hole in the electron beam spatial profile in the direction of the reflected laser beam in the far field. This was shown as a clear signature of the laser-electron beam interaction in vacuum and provided some of the most direct evidence of VLA. This first experiment opened the way for the first time to the investigation of dynamics of free relativistic electrons in ultra-intense laser fields. 3D Simulations of this process are extremely challenging because the reflected electric field carries a large high harmonic content (see Fig. \ref{fig:UHI}) at broad angles and any spurious numerical dispersion induced by FDTD Maxwell solvers will inevitably affect the spatio-temporal phase of harmonic components by deforming the reflected field and therefore significantly impact the properties of the accelerated VLA electrons. 

Effects of numerical dispersion on high harmonics have already been extensively discussed and we demonstrated that our PSATD local implementation can bring up to two orders of magnitude speed-up over FDTD solvers ro reach convergence \cite{Blaclard2016}. Here we will focus on VLA electron properties. Reproducing accurately the features that were observed in the experiment has remained elusive with standard FDTD PIC codes. Hence, the reproduction of the experimental features and the numerical convergence was used as a metric of success for our new pseudo-spectral PIC code.

Thanks to the high performance implementation of the PSATD-local solver in our PIC code, we could for the first time use the PSATD solver at very large scale on over $260k$ cores ($16384$ nodes) on MIRA to benchmark and quantify its huge benefit over standard solvers in terms of time-to-solution and memory for achieving a given precision, as presented in the remainder of this section. Physical parameters/configuration used in simulations were comparable to the experimental ones in \cite{Thevenet2015}: the femtosecond high-intensity laser (intensity $I\approx10^{19}W.cm^{-2}$) reflects at $45^o$ on the plasma mirror and ejects VLA electrons recorded on a detector normal to the specular reflection direction (see central panel on Fig. \ref{fig:UHI}). 3D simulation box dimensions are $50\lambda \times 30\lambda \times 70\lambda$ along $x,y,z$ directions where $\lambda$ is the laser wavelength and $(x,z)$ the plane of incidence of the laser on the target. $16$ plasma pseudo-particles per cell (electrons and ions) were used in 3D. Spatial resolutions was varied from $66$ cells per $\lambda$ to $330$ cells per $\lambda$.   

\subsubsection{Applications and timers} 
\textbf{Warp :}Warp \cite{Warp} is an extensively developed open-source 3D Particle-In-Cell (PIC) code designed to simulate a rich variety of physical processes including laser-plasma interactions at high
laser intensities. Warp is written in a combination of 1) Fortran for efficient implementation of computationally intensive tasks 2) Python for high level specification and control of simulations and 3) C for interfaces between Fortran and Python.

Warp has now been routinely used for many years on NERSC supercomputers\footnote{ MCurie/Seaborg/Bassi/Franklin/Hopper/Edison/Cori}, on Mira at Argonne National Laboratory and other platforms by many scientists worldwide. The last developments of Warp added the advanced ultrahigh-order scalable Maxwell solver described in the preceding section, that is based on domain decomposition with local FFTs. 

\textbf{PICSAR: }Under the auspices of the NERSC Exascale Science Application Program (NESAP), and now DOE's Exascale Project, a full Fortran 90 high-performance PIC library PICSAR (``Particle-In-Cell Scalable Application Ressource'') was recently developed by our team \cite{PICSAR}. This library contains optimized versions of the Warp electromagnetic PIC kernel subroutines. PXR includes numerous optimization strategies to fully benefit from the three levels of parallelisms (Internode, Intranode, Vectorization) offered by current and upcoming architectures (exascale). 

In particular, thanks to the developments made in PXR (some developments are detailed in \cite{Vincenti2017}), Warp+PXR is now a highly optimized code and includes MPI dynamic load balancing at the internode level, optimized MPI stencil communications, hybrid MPI/OpenMP parallelization of the PIC loop, particle tiling and sorting for optimal cache reuse/memory locality and good shared memory OpenMP scaling/intra-node load-balancing, threaded FFTW \cite{FFTWref} for the advanced Maxwell solvers, as well as cutting edge SIMD algorithms for efficient vectorization of hotspots routines \cite{Vincenti2017}. Other optimizations notably include use of MPI-IO for efficient parallel dumping of particles and fields. PXR has been coupled back to Warp through a python layer, by defining a python class that re-defines most of the time consuming Warp methods of the PIC loop. 

PXR has also now been entirely ported to the new Intel KNL architectures and shows very good performances in the early benchmarks done on NERSC's Cori phase 1 and 2. The Warp+PXR simulation tool is now routinely used on NERSC supercomputers in support of laser-plasma experiments performed at LBNL on the BELLA PW laser and also at CEA Saclay in France on the 100 TW laser UHI100. The PIC loop in Warp+PXR can be run as a set of python routines calling Fortran HPC routines using Forthon or as a standalone full Fortran code.  We used the second option for scaling tests on simplified physics problems on Mira and Cori. 

\textbf{Timers: }Timings that are reported in this section were performed with calls to MPI\_WTIME().

\subsubsection{System and environment} 

PXR simulations have been run on two large-scale systems (i) The Mira supercomputer at the Argonne Leadership Computer Facility (ALCF) and (ii) The Cori supercomputer at the National Energy Research Scientific Computing Center (NERSC). 

\textbf{Mira:} Mira \cite{Mira}, an IBM Blue Gene/Q supercomputer at the Argonne Leadership Computing Facility, is equipped with 786,432 cores, 768 terabytes of memory and has a peak performance of 10 petaflops. Mira's 49,152 compute nodes have a PowerPC A2 1600 MHz processor containing 16 cores, each with 4 hardware threads, running at 1.6 GHz, and 16 gigabytes of DDR3 memory. A 17th core is available for the communication library.  IBM's 5D torus interconnect configuration, with 2GB/s chip-to-chip links, connects the nodes, enabling highly efficient computation by reducing the average number of hops and latency between compute nodes. Environment use to compile/link the code: MPICH3, OpenMP 3.0, powerpc-gnu-linux-gcc-cnk v4.4.7 (bgqtoolchain-gcc447), FFTW v3.3.5, Compiler options: '-O3 -fopenmp'. 

\textbf{Cori:} U.S. DOE NERSC's newest supercomputer, named Cori \cite{Cori} and ranked five of the 500 most powerful supercomputers in the world, includes the Haswell partition (Phase I) and the KNL partition (Phase II). We used the KNL partition, which has a (theoretical) peak performance of 27.9 petaflops/sec, 9,688 compute nodes (658,784 cores in total), and 1 PB of memory. 
Each node is a single-socket Intel Xeon Phi Processor 7250 ("Knights Landing") processor with 68 cores per node at 1.4 GHz. Each core has two 512-bit-wide vector processing units. Each core has 4 hardware threads (272 threads total). Two cores form a tile. The peak flops counts are 44 GFlops/core, 3 TFlops/node and 29.1 PFlops total.

Concerning memory, each node has 96 GB DDR4 2400 MHz memory, six 16 GB DIMMs (102 GB/s peak bandwidth), for a total aggregate memory (combined with MCDRAM) of 1 PB. Each node also has 16 GB MCDRAM (multi-channel DRAM), $>$ 460 GB/s peak bandwidth. Each core has its own L1 caches, with 64 KB (32 KB instruction cache, 32 KB data). Each tile (2 cores) shares a 1MB L2 cache. 

The interconnect is a Cray Aries with Dragonfly topology with 45.0 TB/s global peak bisection bandwidth.

The operating system is a lightweigh Linux based on the SuSE Linux Enterprise Server distribution. 
The batch scheduler is SLURM. On Cori, PXR has been compiled using the Intel compiler version 17.0.1.132. 
The MPI implementation is developped by CRAY based on MPICH.
We use a specific option to have memory page size of 2 Mb (huge page) instead of the default 4 Kb page size enabling fastest communications with less fluctuations. 
The code is compiled with the following arguments on KNL: -O3 -xMIC-AVX512 -align array64byte. Libraries/API: FFTW v3.3.5, OpenMP 4.0. 

\subsection{Performance Results} 
\subsubsection{Maxwell solver only}

\begin{figure}[h]
\includegraphics[width=\linewidth]{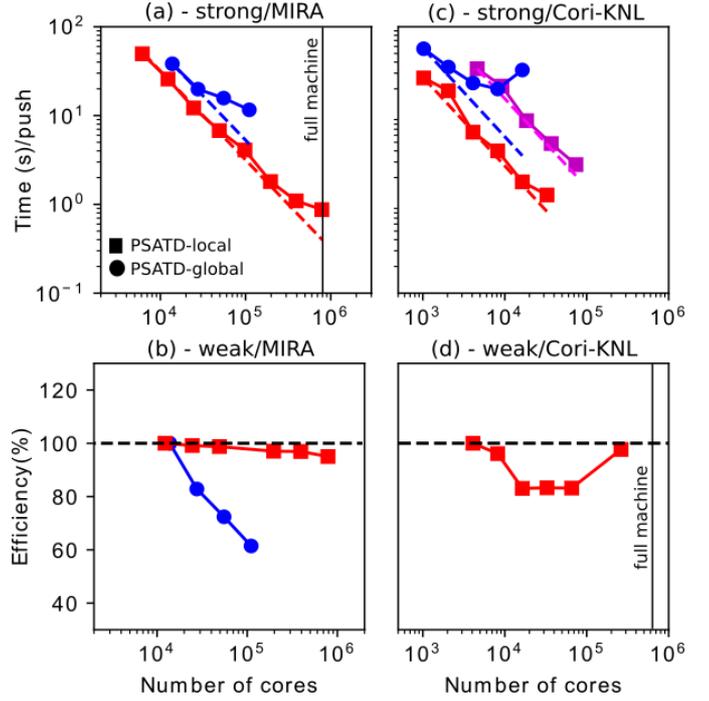}
\caption{\label{fig:scaling_solver_only} \small Strong/Weak scaling of the PSATD Maxwell solver on MIRA/BG-Q and Cori-KNL machines. In each panel, blue dots represent performance data corresponding to the global implementation of the PSATD solver using global FFTs (PSATD-global). Squares represent our new implementation employing local FFTs (PSATD-local). For all tests, we used 8 OpenMP threads per MPI process, a PSATD solver order of $p=100$. For the PSATD-local solver $8$ guard cells were used.  (a) Strong scaling on MIRA on a problem size $1024\times1024\times27648$  (b) Weak scaling on MIRA for a problem size starting at $1024\times1024\times3056$ (c) Strong scaling on Cori-KNL for problem sizes  $512\times512\times110952$ (magenta) and $512\times512\times27648$ (red and blue) (d) weak scaling on Cori-KNL for a problem size starting at $384\times768\times768$. }
\end{figure}

Fig. \ref{fig:scaling_solver_only} presents performance results (weak/strong scaling) of global and local implementations of the pseudo-spectral PSATD Maxwell solver on Cori-KNL and Mira machines. 

 The global implementation (PSATD-global) employs the global 3D distributed FFTs/IFFTs of the FFTW-MPI library (parallelized using MPI. Threading was activated). 

\begin{figure*}[h]
\centering
\includegraphics[width=\linewidth]{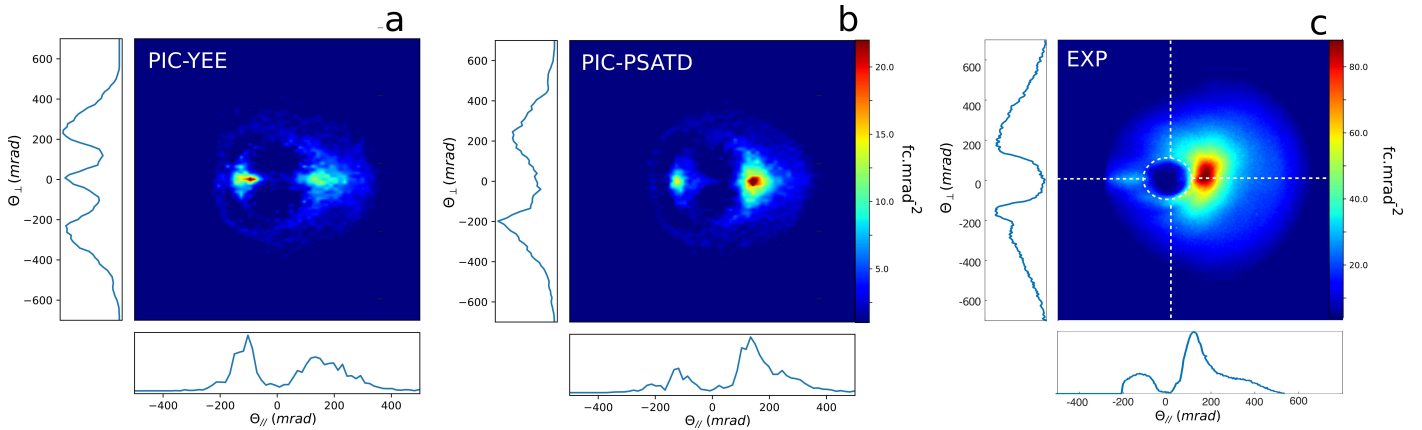}
\caption{\label{fig:fig3}  \small The panels represent the angular distribution of VLA electrons in a plane orthogonal to the direction of specular reflection of the incident laser on the plasma mirror, where $\theta_{\parallelsum}$ is the angle in the incidence plane, $\theta_{\perp}$ the angle in the transverse direction to the incidence plane (see detector on Fig. \ref{fig:UHI}), $\theta_{\parallelsum}=0$ is the specular reflection direction, and $\theta_{\parallelsum}=\pi/4$ rad is the normal to the target. Results are shown for (a) 3D run with the Yee Maxwell solver; (b) 3D run with the PSATD solver at order 100 and with $8$ guard cells;(c) experimental electron distribution obtained from plasma plasma mirror experiments with the UHI100 laser in CEA Saclay \cite{Thevenet2015}. 3D simulations have been run at resolution of  $\approx 60$ cells per laser wavelength $\lambda$.}
\end{figure*}

As detailed in section $3$, our new local implementation (PSATD-local) instead uses a cartesian MPI domain decomposition with local 3D FFTs/IFFTS performed on each subdomain using FFTW threaded version. We used $8$ guard cells at the margin of each MPI-subdomain that are exchanged between neighboring subdomains each time step. For order $p=100$ PSATD (used here), we can demonstrate \cite{Vincenti2016a} that $8$ guard cells are enough to  reduce truncation errors amplitude and their growth to a negligible level in the simulations that were conducted.

The results detailed in Fig. \ref{fig:scaling_solver_only} demonstrate excellent strong and weak scaling of our PSATD-local implementation (red squares on panels (a) and (b)) on up to $800k$ cores on the full MIRA machine ($96\%$ efficiency for the weak scaling on the full machine). As a comparison, the common PSATD-global implementation (blue dots on panels (a) and (b)) performs very poorly at very large scale. On panel (a), the strong scaling efficiency drops at around 20k cores. On panel (b) the weak scaling efficiency drops dramatically with the number of cores at large scale due to the increase in the volume of exchanged data between MPI processes as required by the transposition in global FFTs. For the strong scaling, we could not scale the PSATD-global implementation beyond $120k$ cores because the FFTW implementation only allows CPU split along one dimension in space (last dimension in Fortran). For the weak scaling, we could not obtain scaling data beyond $120k$ cores because FFTW-MPI was requiring too much memory. 

On Cori-KNL, the PSATD-local solver demonstrated excellent strong scaling on up to $100k$ cores (red/magenta curves on panel (c)) and very good weak scaling (red squares on panel (d)) with $98\%$ efficiency on $260k$ cores (half machine). 
On the contrary, the PSATD-global implementation exhibits again very poor performance. The strong scaling efficiency drops at $10k$ nodes and execution time even increases with the number of MPI processes due to MPI exchanges in the global FFT. At $20k$ nodes, we already have almost $2$ orders of magnitude speed-up between the local and global PSATD in terms of time-to-solution.   
Notice that for panels (a) and (c), the PSATD-global (red curve) is slower than the PSATD-local (blue curve) for the same problem size even for a low number of cores. This is due to the fact that the FFT complexity varies as $N \log N$ and the global-FFT is performed on an array size $N$ larger than the one for each individual local FFTs, in addition to performing a global transpose that is not needed with PSATD-local.

\subsubsection{Full physic simulations of plasma mirrors}

Fig. \ref{fig:fig3} shows 3D PIC simulation results of relativistic electrons accelerated by the Vacuum Laser Acceleration mechanism. Here we assess convergence rate of FDTD and PSATD solvers on VLA electron properties.\begin{figure}[h!]
\includegraphics[width=\linewidth]{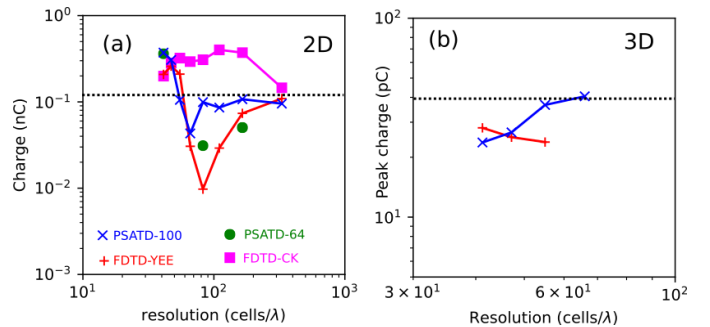}
\caption{\label{fig:fig4} \small Evolution of charge in the VLA beam with resolution (in cells per laser wavelength $\lambda$) from 2D PIC simulation in panels (a). In (a) charge is integrated angularly between target normal and specular direction ($\theta_{\parallelsum}>0$). Panel (c) shows 3D PIC simulation, where charge is angularly integrated in a zone of $\approx 20 mrad\times20 mrad$ around the beam maximum at $\theta_{\parallelsum}\approx 180mrad$ and $\theta_{\perp}\approx -6mrad$ (cf. Fig. \ref{fig:fig3}).}
\end{figure}
 Panels (a) and (b) show angular distribution of VLA electrons obtained with Yee and PSATD solvers at similar resolutions.

One can see that the PSATD solver reproduces all characteristic features of the experimental measurements on panel (c):  the VLA electron beam is located between the target normal and specular direction ($\theta_{\parallelsum}>0$) and there is a clear hole in the electron distribution projected along $\theta_{\perp}$ and $\theta_{\parallelsum}$. On the contrary the simulation performed with the Yee solver leads to wrong results (electrons located in the hole, much more electrons below the specular direction i.e for $\theta_{\parallelsum}<0$). These spurious artefacts are induced by numerical dispersion of the Yee scheme.

 Convergence tests were performed at different resolutions (cf. Fig. \ref{fig:fig4}) in 2D and 3D.  Fig. \ref{fig:fig3} (b-c) and  demonstrate convergence of the PSATD at order $100$ already for $66$ cells per laser wavelength $\lambda$ (max resolution carried out in 3D), while Yee and CK solvers would at least require $\approx300$ cells$/\lambda$ (cf. Fig. \ref{fig:fig4} (a)) . 

\begin{table}[h]
\small
\centering
\caption{Resources to solution in 3D}\label{wrap-tab:1}
\begin{tabular}{ccc}\\\toprule  
\textbf{Solver} & \textbf{Resolution} & \textbf{core$\times$hours} \\ 
& (cells/$\lambda$)&  \\ \midrule
PSATD ($p=100$)  & 66    & 7M\\  
Yee (estimated)        &  300 & 3000M\\  
CK  (estimated)       &  300  & 1700M\\  \bottomrule
\end{tabular}
\end{table}

In addition, we also observe that PSATD at order 64 converges more slowly than at order 100, justifying the use of ultrahigh order. These simulation results thus enable real measurement (PSATD solver) and estimates (Yee,CK) of resources-to-solution needed for convergence, which are given in table \ref{wrap-tab:1}, with speedups ranging between $240\times$ to $430\times$ for PSATD (order $100$) over standard FDTD solvers. Note that we provide estimates for Yee and CK solvers based on projections from 2D simulations of Fig. \ref{fig:fig4} (a), as we could not carry 3D simulations on up to $\approx300$ cells$/\lambda$ on even the largest available machines at time of writing.

\section{Future implications} 
These results may have a huge fundamental impact on UHI physics, as 3D accurate PIC simulations are essential to the detailed understanding of these new laser-plasma interaction regimes, where particle motion is highly relativistic and very short-wavelength radiations can be emitted at broad angles. In particular, by accurately capturing  the spatio-temporal properties of ultra-compact attosecond electron and X-UV light sources from plasma mirrors, the new method can be used to identify optimal regimes of productions of these sources for performing promising application experiments such as attosecond pump-probe experiments or time-resolved diffraction imagery.

 Beyond plasma mirrors, our general approach may also impact other applications relying on the self-consistent modeling of plasmas or charged particle beam. For instance, by eliminating dispersion and minimizing heating errors, the algorithm proposed here is also especially important to the conception of next generation PW laser wakefield acceleration experiments where the orders of magnitude higher electron beam quality (in emittance and/or energy spread) that are required will necessitate a similar increase in numerical accuracy. In fact, the new algorithm is one of the key innovations that are at the heart of DOE's Exascale application project ``Exascale modeling of advanced particle accelerators'' \cite{ECP}. Another innovation, that is key to the efficient modeling of laser-plasma accelerators, is the reduction of the number of time steps that are required to model the propagation of a laser beam through an under-dense plasma by orders of magnitude, by choosing an optimal relativistic Lorentz boosted frame of reference for the calculation \cite{Vayprl07}. As it turns out, the PSATD solver is paramount to a new method \cite{Lehe2016,KirchenPOP2016} eliminating the so-called ``numerical Cherenkov instability'' \cite{Godfreyjcp74}, and thus enabling the orders of magnitude speedup of the Lorentz boosted frame method to its full potential.

In addition, by reducing the number of space and time steps required for a given accuracy to the solution of a plasma physics problem, the new method can considerably reduce the time to solution in both the design of devices and the study of fundamental science in the area of plasma and electro-energetic physics, including -but not limited to- laser plasma acceleration and relativistic optics (e.g. filamentation, high harmonic generation, ion acceleration). As we demonstrated in this paper, this renders for the first time converged 3D simulations of plasma mirrors accessible on existing supercomputers and 2D simulations on a local workstation. 

Finally, the method presented here has the potential for having a broader impact on a large class of computational physics problems, as the underlying concept is applicable in principle to various initial value problems that can be treated by FFT-based pseudo-spectral methods. The integration of Maxwell's equations is just one example of the so-called `initial value' problems that are some of the most important in science, where differential equations are integrated numerically in time based on the initial state of the system and time-dependent source terms. Examples of initial value problems include the diffusion equation, Vlasov equation, general relativity, Schrödinger equation, etc (indeed our team successfully tested the method on the modeling of the heat equation). When these problems can be treated by FFT-based pseudo-spectral methods, the new parallelization method demonstrated here may in some cases apply and enable better scalability, especially on future exascale supercomputers and beyond, where scaling to ultrahigh concurrency will be paramount.

 \section*{Acknowledgement}

We thank Guillaume Blaclard, Dr. Remi Lehe, Dr. Brendan Godfrey, Dr. Irving Haber and Dr. Fabien Quere for fruitful discussions. We are also very grateful to Dr. A. Leblanc who provided us with the experimental results performed on the UHI100 laser at CEA Saclay. This work was supported by the European Commission through the Marie Sk\l owdoska-Curie actions (Marie Curie IOF fellowship PICSSAR grant number 624543) as well as by the Director, Office of Science, Office of High Energy Physics, U.S. Dept. of Energy under Contract No. DE-AC02-05CH11231, the US-DOE SciDAC program ComPASS, and the US-DOE program CAMPA. An award of computer time (PICSSAR\_INCITE) was provided by the Innovative and Novel Computational Impact on Theory and Experiment (INCITE) program. This research used resources of the Argonne Leadership Computing Facility, which is a DOE Office of Science User Facility supported under Contract DE-AC02-06CH11357 as well as resources of the National Energy Research Scientific Computing Center, a DOE Office of Science User Facility supported by the Office of Science of the U.S. Department of Energy under Contract No. DE-AC02-05CH11231.

This document was prepared as an account of work sponsored in part
by the United States Government. While this document is believed to
contain correct information, neither the United States Government
nor any agency thereof, nor The Regents of the University of California,
nor any of their employees, nor the authors makes any warranty, express
or implied, or assumes any legal responsibility for the accuracy,
completeness, or usefulness of any information, apparatus, product,
or process disclosed, or represents that its use would not infringe
privately owned rights. Reference herein to any specific commercial
product, process, or service by its trade name, trademark, manufacturer,
or otherwise, does not necessarily constitute or imply its endorsement,
recommendation, or favoring by the United States Government or any
agency thereof, or The Regents of the University of California. The
views and opinions of authors expressed herein do not necessarily
state or reflect those of the United States Government or any agency
thereof or The Regents of the University of California.

\bibliographystyle{unsrt}




\end{document}